\documentclass[twocolumn]{aastex7}

\shorttitle{Helium in GJ 3090b}
\shortauthors{Vissapragada et al.}

\begin{document}

\correspondingauthor{Shreyas~Vissapragada}

\title{WINERED Detects a Strong Atmospheric Outflow on the Sub-Neptune GJ 3090b\footnote{This paper includes data gathered with the 6.5 meter Magellan Telescopes located at Las Campanas Observatory, Chile.}}

\author[0000-0003-2527-1475]{Shreyas~Vissapragada} 
\affiliation{Carnegie Science Observatories, 813 Santa Barbara Street, Pasadena, CA 91101, USA}
\email[show]{svissapragada@carnegiescience.edu}

\author[0000-0003-0525-9647]{Zifan~Lin} 
\affiliation{Department of Earth, Atmospheric and Planetary Sciences, Massachusetts Institute of Technology, 77 Massachusetts Avenue, Cambridge, MA 02139, USA}
\affiliation{Department of Physics and McDonnell Center for the Space Sciences, Washington University, St. Louis, MO 63130, USA}
\email{lzifan@wustl.edu}

\author[0000-0002-7500-7173]{Annabella~Meech} 
\affiliation{Space Telescope Science Institute, 3700 San Martin Drive, Baltimore, MD 21218, USA}
\email{ameech@stsci.edu}

\author[0000-0002-8466-5469]{Collin~Cherubim} 
\altaffiliation{NHFP Sagan Fellow}
\affiliation{Department of Astronomy \& Astrophysics, University of Chicago, Chicago, IL 60637, USA}
\email{collinc@uchicago.edu}

\author[0000-0002-2248-3838]{Leonardo~A.~Dos~Santos} 
\affiliation{Space Telescope Science Institute, 3700 San Martin Drive, Baltimore, MD 21218, USA}
\email{ldsantos@stsci.edu}


\author[0000-0002-5812-3236]{Aaron~Householder}
\affiliation{Department of Earth, Atmospheric and Planetary Sciences, Massachusetts Institute of Technology, 77 Massachusetts Avenue, Cambridge, MA 02139, USA}
\affil{Kavli Institute for Astrophysics and Space Research, Massachusetts Institute of Technology, Cambridge, MA 02139, USA}
\email{aaron593@mit.edu}

\author[0009-0005-4890-3326]{Krishna~Kanumalla}
\affiliation{School of Earth and Space Exploration, Arizona State University, 781 Terrace Mall, Tempe, AZ 85287, USA}
\email{skanuma1@asu.edu}

\author[0000-0003-3204-8183]{Mercedes L\'{o}pez-Morales}
\affiliation{Space Telescope Science Institute, 3700 San Martin Drive, Baltimore, MD 21218, USA}
\email{mlopez-morales@stsci.edu}

\author[0000-0002-0765-431X]{Andrew~McWilliam}
\affiliation{Carnegie Science Observatories, 813 Santa Barbara Street, Pasadena, CA 91101, USA}
\email{andy.ociw@gmail.com}

\author[0000-0001-6315-7118]{William~Misener}
\affiliation{Carnegie Science Earth \& Planets Laboratory, 5241 Broad Branch Road, NW, Washington, DC 20015, USA}
\email{wmisener@carnegiescience.edu}

\author[0009-0000-3527-8860]{Ava~Morrissey}
\affiliation{Carnegie Science Observatories, 813 Santa Barbara Street, Pasadena, CA 91101, USA}
\email{amorrissey@carnegiescience.edu}

\author[0000-0001-9518-9691]{Morgan~Saidel}
\affiliation{Carnegie Science Observatories, 813 Santa Barbara Street, Pasadena, CA 91101, USA}
\email{msaidel@caltech.edu}

\author[0000-0002-5547-3775]{Jessica~J.~Spake} 
\affiliation{Carnegie Science Observatories, 813 Santa Barbara Street, Pasadena, CA 91101, USA}
\email{jspake@carnegiescience.edu}

\author[0009-0008-2801-5040]{Johanna~Teske}
\affiliation{Carnegie Science Earth \& Planets Laboratory, 5241 Broad Branch Road, NW, Washington, DC 20015, USA}
\affiliation{Carnegie Science Observatories, 813 Santa Barbara Street, Pasadena, CA 91101, USA}
\email{jteske@carnegiescience.edu}

\author[0000-0003-0354-0187]{Nicole~L.~Wallack}
\affiliation{Carnegie Science Earth \& Planets Laboratory, 5241 Broad Branch Road, NW, Washington, DC 20015, USA}
\email{nwallack@carnegiescience.edu}

\begin{abstract}
Sub-Neptunes are the most common short-period ($P<100$~d) planets known. A major goal for the field is to build a comprehensive understanding of their compositions and evolutionary histories, and an important tool for achieving this goal is metastable helium (He$^*$) transmission spectroscopy, which traces atmospheric escape. In this paper, we report the detection of He$^*$ in the atmosphere of the sub-Neptune GJ 3090b at high spectral resolution. We observed two consecutive transits of this planet with Magellan II/WINERED, and detected peak He$^*$ absorptions during transit of $2.033\pm0.086\%$ (23.6$\sigma$) and $1.619\pm0.122\%$ (13.3$\sigma$), respectively. The first transit was affected by a stellar flare near egress while the second transit was unaffected. The signal is variable in amplitude and deeper than anticipated by the JWST/NIRISS observations of \citet{Ahrer2025}. We do not observe a significant Doppler shift, and the signal has a FWHM of $25$~km~s$^{-1}$, implicating photoevaporation. Our outflow modeling suggests that $Z_\mathrm{atm}\lesssim100\times$ solar, as the observed He$^*$ amplitudes are challenging to achieve in a metal-rich atmosphere. To match the muted transmission spectra from JWST and VLT/CRIRES+, GJ 3090b's lower atmosphere is likely covered by aerosols. Finally, we found that GJ 3090b has an atmospheric lifetime of $\sim300$~Myr, much shorter than previously reported. This adolescent ($\sim1$~Gyr) sub-Neptune is currently in a transformative phase of photoevaporative evolution. 
\end{abstract}

\section{Introduction} \label{sec:intro}
More than half of the stars in our Galaxy host a short-period ($P < 100$~d) planet between the sizes of Earth and Neptune ($R_\Earth < R_p < 4R_\Earth$) \citep{Howard2012, Dressing2013, Bean2021}. The distribution of planetary radii in this size range is bimodal \citep{Fulton2017}: smaller, ostensibly rocky planets with $R_p\lesssim1.5R_\Earth$ (``super-Earths'') are separated from larger $2R_\Earth \lesssim R_p \lesssim 4R_\Earth$ ``sub-Neptunes'' by a radius valley between roughly $1.5R_\Earth$ and $2R_\Earth$. The structures, compositions, and evolutionary histories of these remarkably common small planets are important open problems in exoplanet astronomy as they are challenging to infer from mass and radius alone \citep[e.g.][]{Zeng2019, Luque2022, Teske2025}. 

Transmission spectroscopy can provide insight into the structures and compositions of small exoplanets. Sub-Neptune transmission spectra from JWST have already revealed a broad range of atmospheric compositions, from H$_2$/He-dominated atmospheres on K2-18b and TOI-421b \citep{Madhusudhan2023, Davenport2025} to a mixture of H$_2$/He/H$_2$O on TOI-270d \citep{Benneke2024} to an H$_2$O-dominated atmosphere on GJ 9827d \citep{Piaulet-Ghorayeb2024}. The range of mean molecular weights (MMWs) may result from broad differences in planetary structure. The lower-MMW atmospheres match the canonical expectation for a rocky core enveloped by H$_2$/He, while the higher-MMW atmospheres may be suggestive of water-rich bulk compositions, i.e. the ``water world'' scenario \citep{Adams2008, Rogers2011, Bitsch2021, Luque2022, Chakrabarty2026}. Fractionating atmospheric loss \citep{Cherubim2025, Louca2025, Valatsou2026} and geochemical interactions between the core and the atmosphere \citep{Schlichting2022, Misener2023, Gupta2025, Heng2025} may also affect the observed compositions. A larger sample size would help to distinguish between some of these possibilities, but most sub-Neptune transmission spectra are featureless, which could either be due to high MMWs, high-altitude aerosols, or both \citep[e.g.][]{Wallack2024, Teske2025}.

Planetary evolution in the 1$R_\Earth < R_p < 4R_\Earth$ regime also remains debated. The radius valley was predicted by thermal escape models assuming H$_2$/He envelopes atop rocky cores \citep{Lopez2013, Owen2013}, but different thermal escape mechanisms can explain the data equally well, including XUV-driven photoevaporation \citep{Owen2017} and core-powered mass loss \citep{Ginzburg2018, Gupta2020}. The relative contributions of these mechanisms are challenging to distinguish using population statistics alone \citep{Rogers2021}, and they may sculpt the small planet population together rather than operating in isolation \citep{Owen2024, Misener2026}. Moreover, multiple groups have argued that the valley need not be fully evolutionary; it could result from fundamental limits to gas accretion for small cores \citep{Lee2021, Lee2022, Nielsen2025} or from the inward migration of ``water worlds'' \citep{Burn2024, Venturini2024}. The abundance of theoretical interpretations for the radius valley highlights a need for additional empirical constraints on sub-Neptune evolution. 

We can make progress on illuminating fundamental small planet properties by studying their upper atmospheres using the metastable He 10833~\AA~(He$^*$) triplet \citep{Nortmann2018, Spake2018}. He$^*$ traces atmospheric escape \citep{Oklopcic2018, Ballabio2025}, which makes it a valuable probe of planetary evolution. For example, the He$^*$ line shape is sensitive to outflow velocities and can be constrained precisely \citep{Zhang2023a}, so He$^*$ observations can directly distinguish between photoevaporation \citep[for which flow speeds are typically faster, $\gtrsim10$~km~s$^{-1}$;][]{Murray-Clay2009, Owen2023} and core-powered mass-loss \citep[for which flow speeds are typically slower, $\sim2-3$~km~s$^{-1}$;][]{Ginzburg2018, Misener2025}. Additionally, He$^*$ is less likely to be detected in high-MMW atmospheres ($Z_\mathrm{atm}\gtrsim100Z_\Sun$), like those expected in the water-world scenario \citep{Rogers2026}. Such atmospheres are predicted to have small scale heights, low abundances of helium, and much weaker outflows due to metal and/or molecular cooling \citep{Salz2016, Vissapragada2024, GarciaMunoz2025, Yoshida2025, Zhang2025, Kubyshkina2026, Taylor2026}. He$^*$ detections can therefore help break degeneracies in the interpretation of featureless transmission spectra. 

With these applications of He$^*$ in mind, we initiated the WINERED Helium Consortium, which aims to constrain He$^*$ absorption in a large sample of sub-Neptune-sized exoplanets. WINERED \citep{Ikeda2022, Otsubo2024} is a high-resolution ($R\sim68,000$) near-infrared spectrograph mounted on Magellan-II/Clay and has already demonstrated good precision on-sky for He$^*$ transit spectroscopy \citep{Vissapragada2024, Cherubim2026}. Our goal is to constrain He$^*$ absorption (or lack thereof) to better than 0.2\% precision in the line core for $\gtrsim30$ targets. We selected this precision goal because most sub-Neptune He$^*$ detections to date have been around $\sim$1\% in amplitude \citep{Orell-Miquel2024}, and 0.2\% precision would correspond to a 5$\sigma$ detection in this case. 

In this paper, we report a high-resolution detection of strong He$^*$ absorption on GJ 3090b, a relatively low-density sub-Neptune ($M_p = 4.5\pm0.5M_\Earth$, $R_p = 2.2\pm0.1R_\Earth$, $\rho_p = 2.4\pm0.3$~g~cm$^{-3}$) on a $P = 2.85$ day orbit around an M2V star \citep[giving it an equilibrium temperature of $T_\mathrm{eq} = 723\pm13$~K;][]{Almenara2022, Lamontagne2026}. The system is young; \citet{Almenara2022} estimate a gyrochronological age of 1.02$^{+0.23}_{-0.15}$ Gyr. GJ 3090b is also one of the best-known sub-Neptunes for transmission spectroscopy, with a Transmission Spectroscopy Metric \citep{Kempton2018} of 182$^{+27}_{-23}$. Previous JWST/NIRISS observations revealed escaping helium in its atmosphere \citep{Ahrer2025}, but neither JWST nor ground-based VLT/CRIRES+ observations could confidently detect molecular absorption from the lower atmosphere \citep{Parker2025}. In Section~\ref{sec:observations}, we describe our observations and subsequent data reduction process, and in Section~\ref{sec:results}, we describe the results of the observations. We compare the He$^*$ absorption signal to models in Section~\ref{sec:modeling} to get a sense for the planetary outflow physics, and we conclude and discuss the broader implications of our results in Section~\ref{sec:conclusion}.

\section{Observations and Data Reduction}
\label{sec:observations}

\begin{figure*}[ht!]
    \centering
    \includegraphics[width=\textwidth]{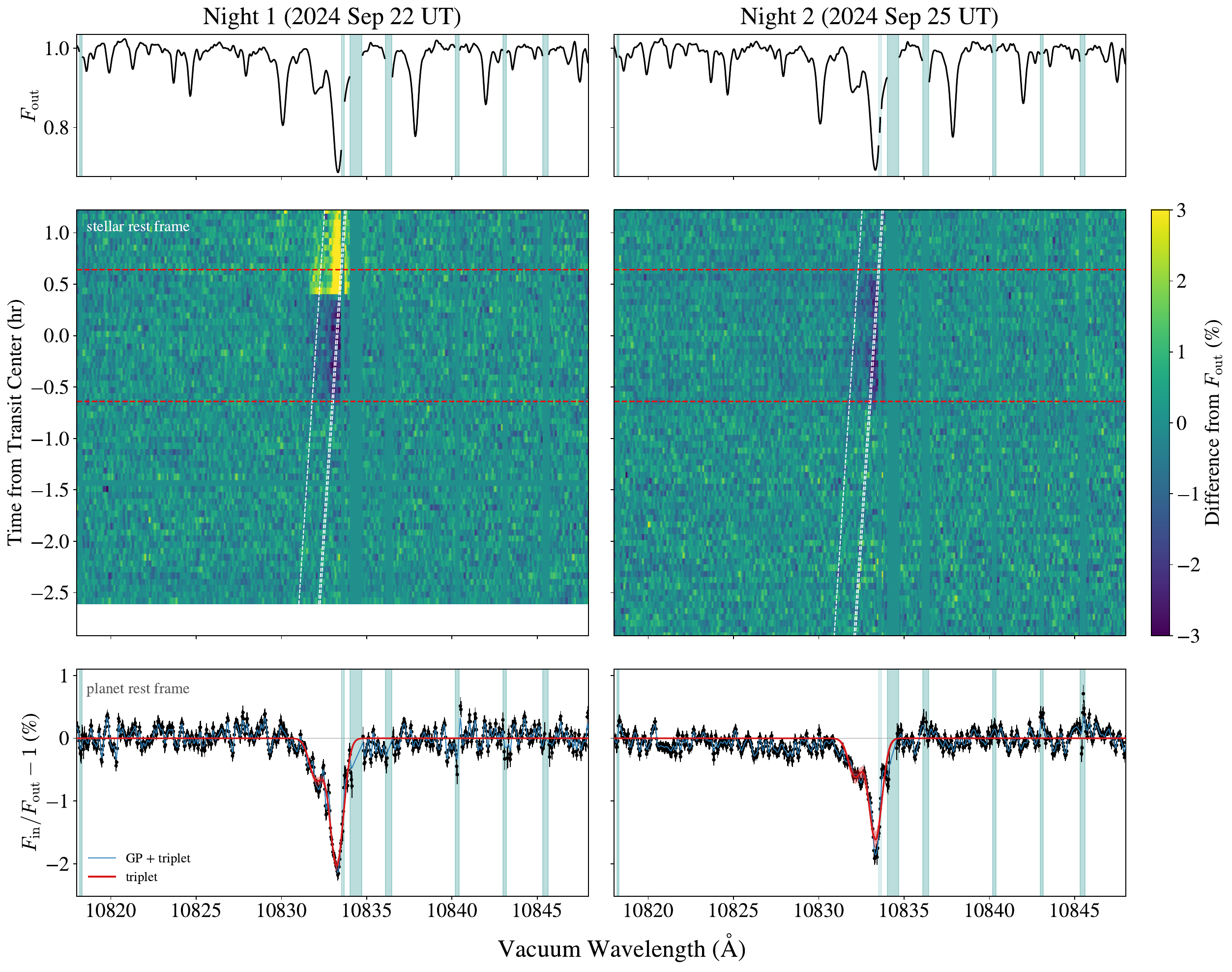}
    \caption{Timeseries spectroscopy of GJ 3090b on the first night (left) and second night (right). The top panels show the out-of-transit template spectra for each night, the middle panels show the timeseries spectra in the stellar rest frame (where colors indicate percentage difference from the night's out-of-transit template), and the bottom panels show the transmission spectra in the planetary rest frame along with the best-fit three-Gaussian model in red (with shaded 1$\sigma$ uncertainty interval) and the best-fit GP instrumental noise model in blue. The transmission spectra are shown at pixel resolution ($R\sim300,000$), so correlated noise is expected. Note that the first night is clearly affected by a flare starting just before egress; the flare-affected data were not used in the analysis, and are shown for completeness. Red dashed lines indicate the predicted beginning and end of transit on each night, and the diagonal white lines indicate the positions of the helium triplet lines following the planetary rest frame. The masks for the tellurics are shown with the teal vertical bars.}
    \label{fig:timeseries}
\end{figure*}

We observed two consecutive transits of GJ 3090b on 2024 September 22 and 2024 September 25 with Magellan II/WINERED \citep{Ikeda2022, Otsubo2024}. For both observations, we operated WINERED in HIRES-Y mode with the 100~$\mu$m slit to maximize resolving power. We took continuous 180~s exposures with an ABBA nodding strategy (with a nodding amplitude of 5\arcsec) to subtract the telluric OH emission lines and to mitigate the effects of persistence between exposures. Our observations lasted from 06:08 UT to 10:00 UT on the first night (airmass 1.05 to 1.54 with minimum airmass 1.05) and from 01:40 UT to 06:29 UT on the second night (airmass 1.06 to 1.57 with  minimum airmass 1.05). The predicted mid-transit times \citep[using ephemerides from][]{Almenara2022} were 08:51 UT and 05:20 UT, respectively, with uncertainties less than 3~min, so the $T_\mathrm{14} = 1.28$~hr transit events were covered comfortably with ample baseline. In all, 64 spectra were obtained on the first night, and 80 spectra were obtained on the second night.

We then reduced the data using the method from \citet{Vissapragada2024} with some updates that we summarize briefly here. We used the public \textsf{WARP} pipeline \citep{Hamano2024} to obtain a one-dimensional spectrum (with wavelengths in vacuum) for each exposure in order 163, which contains He$^*$. We removed one spectrum with low S/N on the first night. We also noticed that the detector persistence had not yet settled at the beginning of the second night of observations, so we dropped the first 10 spectra for that night to avoid biasing the out-of-transit template. Excluding these frames, the median per-pixel S/N was 142.4 on the first night and 146.3 on the second night.

WINERED spectra exhibit a systematic flux difference between spectra taken at the A and B positions, which is often clearly seen around 10810~\AA~when using the nominal nod positions \citep[e.g.][]{Cherubim2026}. To estimate the systematic, we divided each spectrum by its median, and then took the ratio of the median-normalized A and B spectra for each AB and BA pair. We then averaged the systematic spectra obtained for all AB and BA pairs to obtain a higher S/N estimate of the effect. Finally, we multiplied all spectra taken at the B position by the estimate to correct the systematic offset. This correction removed variations correlated with nod position in the final timeseries spectrum.

We then continuum-normalized the spectra. We first constructed a line mask, starting by averaging together all median-normalized spectra from a given night and masking a broad region around the helium line from 10830~\AA\,\,to 10835~\AA. We then iteratively fit a third-order Chebyshev polynomial to the combined spectrum, calculated the median absolute deviation (MAD) from the polynomial residuals, and added pixels more than $1\times$MAD below the polynomial to the line mask. This procedure was iterated 11 times to generate a global line mask for each night. Then, we applied the global line mask to each individual spectrum and fit a third-order Chebyshev polynomial to the line-free regions. Next, we refined the wavelength solution by fitting for a shift between the observed positions of the telluric H$_2$O absorption lines and model telluric lines \citep[using the HITRAN interface \textsf{hapi};][]{Kochanov2016, Gordon2022}. The wavelength shifts were typically $\lesssim0.02$~\AA~on the first night and $\lesssim0.03$~\AA~on the second night. Once the wavelength solutions were refined, we masked telluric H$_2$O lines and any residuals from OH sky subtraction before shifting the data into the stellar rest frame. We then constructed a stellar template using out-of-transit exposures (selected with the \citealt{Almenara2022} ephemerides).

In Figure~\ref{fig:timeseries}, we show the stellar template for each transit in the upper panels. We then divided all of the spectra by the appropriate stellar templates to obtain the normalized timeseries spectra shown in the middle panels of Figure~\ref{fig:timeseries}. We observed a flare that overlapped with egress during the first transit (Appendix~\ref{app:flare}), so we masked out all flare-affected data (the final 17 spectra) when constructing the out-of-transit template and planetary transmission spectrum for the first night. We shifted all of the in-transit spectra into the planetary rest frame using the known orbital properties of GJ 3090b \citep{Lamontagne2026}. We took the average of the in-transit spectra and divided by the out-of-transit template to obtain the planetary transmission spectrum for each night, which are shown in the bottom panel of Figure~\ref{fig:timeseries}. 

\begin{deluxetable*}{lllccc}[t]
\tablewidth{0pt}
\tablecaption{Three-Gaussian Fit Parameters for GJ 3090 b \label{tab:fit}}
\tablehead{
\colhead{Parameter} & \colhead{Variable} & \colhead{Units} & \colhead{Prior} &
\colhead{Posterior (Night 1)} & \colhead{Posterior (Night 2)}}
\startdata
First component amplitude & $A_1$ & \% & $\mathcal{U}(0,\,5)$ & $0.666^{+0.081}_{-0.080}$ & $0.609^{+0.110}_{-0.112}$ \\
Second component amplitude & $A_2$ & \% & $\mathcal{U}(0,\,5)$ & $1.08^{+0.64}_{-0.70}$ & $0.85^{+0.53}_{-0.56}$ \\
Third component amplitude & $A_3$ & \% & $\mathcal{U}(0,\,5)$ & $0.95^{+0.71}_{-0.63}$ & $0.76^{+0.54}_{-0.52}$ \\
Full-width at half-maximum & $\mathrm{FWHM}$ & km\,s$^{-1}$ & $\mathcal{U}(1,\,60)$ & $25.9^{+1.4}_{-1.3}$ & $24.9^{+2.5}_{-2.2}$ \\
Doppler shift & $v_\mathrm{shift}$ & km\,s$^{-1}$ & $\mathcal{U}(-20,\,20)$ & $-0.5^{+1.0}_{-1.0}$ & $2.0^{+1.2}_{-1.2}$ \\
GP amplitude & $\sigma_\mathrm{GP}$ & \% & $\log\mathcal{U}(10^{-4},\,2)$ & $0.144^{+0.008}_{-0.007}$ & $0.138^{+0.009}_{-0.008}$ \\
GP length scale & $\rho_\mathrm{GP}$ & \AA & $\mathcal{U}(0.02,\,3)$ & $0.396^{+0.020}_{-0.019}$ & $0.571^{+0.047}_{-0.042}$ \\
GP damping time scale & $\tau_\mathrm{GP}$ & \AA & $\log\mathcal{U}(0.05,\,50)$ & $0.123^{+0.017}_{-0.015}$ & $0.099^{+0.017}_{-0.015}$ \\
Peak absorption & $A_\mathrm{peak}$ & \% & \nodata & $2.033^{+0.086}_{-0.086}$ & $1.619^{+0.122}_{-0.122}$ \\
Line depth ratio & $R_\mathrm{depth}$ & \nodata & \nodata & $2.95^{+0.36}_{-0.30}$ & $2.59^{+0.54}_{-0.39}$ \\
\enddata
\tablecomments{Peak absorption and line depth ratio are derived rather than fitted parameters.}
\end{deluxetable*}

\section{Results} \label{sec:results}
\subsection{Three-Gaussian Model} \label{sec:gaussian}
Excess absorption at 10833~\AA~is clearly visible in Figure~\ref{fig:timeseries}. We first fit the triplet feature using a three-Gaussian model. We fit for a free non-negative amplitude in each component ($A_1, A_2, A_3$), a common line width for all three components (expressed as a FWHM in km~s$^{-1}$), and a common Doppler shift (relative to the National Institute of Standards and Technology line positions) for all three components ($v_\mathrm{shift}$, also expressed in km~s$^{-1}$). Because the two red lines in the triplet are blended, their posteriors are highly correlated, so we also track and report the maximum absorption in the model. We additionally track the line depth ratio between the maximum absorption and the weaker blue peak, which should be about 8 for an optically thin gas \citep[set by the 1:3:5 degeneracy ratio; e.g.][]{Salz2018}.

We fit this three-Gaussian model to the data using \textsf{emcee} \citep{ForemanMackey2013}. For the noise model, we use a single \textsf{SHOTerm} Gaussian Process (GP) from \textsf{celerite2} \citep{ForemanMackey2017, ForemanMackey2018} with a free amplitude $\sigma_{\rm GP}$, lengthscale $\rho_{\rm GP}$, and damping lengthscale $\tau_{\rm GP}$. A correlated noise model was necessary because the reduced transmission spectra are at roughly pixel resolution, and there are roughly 4.4~pixels per resolution element, so neighboring points are correlated by construction. We sample the model using 36 walkers, running each for 2,000 ``burn-in'' steps before taking 10,000 posterior draws. We ran for longer than 100 integrated autocorrelation times for all sampled parameters, indicating good convergence. We repeated the fits using a Mat\'{e}rn~3/2 kernel for the GP and found that all of our inferences were robust ($<1\sigma$) to the choice of selected GP kernel. The fit priors and posteriors for the nominal SHOTerm GP fit are given in Table~\ref{tab:fit}, and we show the fit in Figure~\ref{fig:timeseries}. We make the following inferences from this simple modeling exercise:

\begin{enumerate}
    \item The outflow is detected at high significance in both transits: on the first night, we detect a peak absorption of $2.033\pm0.086\%$ (23.6$\sigma$ detection significance) and on the second night, we detect a peak absorption of $1.619\pm0.122\%$ (13.3$\sigma$). These are much larger than the 0.5-1\% absorption amplitudes that have been typically observed for sub-Neptunes \citep{Orell-Miquel2024, Zhang2025}, although GJ 1214b may be an exception (\citealt{Orell-Miquel2022}, but see also \citealt{Kasper2020, Spake2022}). The excess absorptions correspond to equivalent radius excess values of $\delta R_p= 2.9R_p$ and $\delta R_p= 2.5R_p$ for an opaque absorbing layer of helium. The Roche lobe extends roughly to $R_\mathrm{Roche}\approx a\big(\frac{M_p}{3M_\star}\big)^{1/3}\approx7.0R_p$, so the total opaque radius probed ($R_p + \delta R_p \approx 3.9R_p$) is over 50\% of $R_\mathrm{Roche}$. This is larger than many low-density gas giants known to be undergoing strong helium escape \citep[e.g. WASP-69b, WASP-107b;][]{Allart2023}. 
    
    \item The He$^*$ transits are far too deep to result from the transit light source effect (i.e. transiting over an inhomogeneous stellar disk) for a 2$R_\Earth$ planet \citep[e.g.][]{Cauley2018, Rackham2018}. If the transit chord were entirely free of He$^*$, the resulting pseudo-signal would be at most $\frac{\delta d}{(1-d)(1-\delta)} = 0.06\%$, where $\delta = 0.14\%$ is the white-light transit depth and $d\approx0.31$ is the disk-integrated He$^*$ line depth we measure from the out-of-transit templates (see top panels of Figure~\ref{fig:timeseries}). This is more than an order of magnitude smaller than the observed signals. A similar argument was made for the sub-Neptune TOI-560b by \citet{Zhang2022}.
    
    \item Although the two measurements cover consecutive transits, the outflow signal appears to be variable even on this short timescale. The difference in peak absorption between the measurements is $0.414\pm0.149\%$, a 2.8$\sigma$ tension. The flare observed during the first transit suggests that short-timescale variations in irradiating flux could help drive the differences in observed He$^*$ absorption \citep{Wang2021, Mercier2025}.
    
    \item The outflow is photoevaporative in the region probed by our observations. The He$^*$ features are broad, with FWHMs of $25.9^{+1.4}_{-1.3}$~km~s$^{-1}$ on the first night and $24.9^{+2.5}_{-2.2}$~km~s$^{-1}$ on the second night. The difference between the two broadening measurements is not significant. The strong broadening is characteristic of photoevaporative flows (for which $v\gtrsim10$~km~s$^{-1}$ is expected), but not of core-powered mass loss \citep[for which $v\lesssim5$~km~s$^{-1}$ is expected; e.g.][]{Owen2023}, and is consistent with most helium observations of sub-Neptunes \citep[e.g.][]{Zhang2023a}. \citet{Misener2026} predict exactly this behavior for GJ 3090b. They find that as a $4.5M_\Earth$, $723$~K planet like GJ 3090b contracts over its evolution, the planet should switch from a bolometric to a photoevaporative outflow at around $5.4R_\Earth$. This is much larger than the observed 2.2$R_\Earth$, suggesting that the planet is well into its photoevaporative outflow phase, in agreement with the strong broadening that we observe.
    
    \item The outflow is not strongly Doppler-shifted. The observation is best fit with shifts of $-0.5^{+1.0}_{-1.0}$~km~s$^{-1}$ on the first night and $+2.0^{+1.2}_{-1.2}$~km~s$^{-1}$ on the second night. While some sub-Neptunes exhibit strong He$^*$ blueshifts indicative of interactions with the stellar wind \citep[e.g.][]{Zhang2023a}, GJ 3090b adds to the growing number of sub-Neptunes without clear evidence for these interactions, including TOI-560b \citep{Zhang2022}, TOI-2134b \citep{Zhang2023}, and LHS 1140b \citep{Cherubim2026}. While the diversity of Doppler shifts could indicate differing wind-wind interactions across these systems, this interpretation is degenerate with apparent Doppler shifts due to an eccentric orbit \citep[e.g.][]{Zhang2023}. Eccentricities are challenging to constrain for small planets. For GJ 3090b, modest eccentricities are still permitted by the RV and JWST transit datasets so long as $e<0.25$ \citep{Ahrer2025, Lamontagne2026}, leaving this as an important factor in determining the expected Doppler shift of the signal.

    \item The observation probes gas with significant optical depth. We clearly detect the weaker blue line on both nights, and detect a line depth ratio of $2.95^{+0.36}_{-0.30}$ on the first night and $2.59^{+0.54}_{-0.39}$ on the second night. These values are significantly discrepant (at 14.0$\sigma$ and 10.0$\sigma$ confidence, respectively) with the expected optical depth ratio of 8 for an optically-thin gas. The line depth ratio is consistent across the two visits. 
\end{enumerate}

\begin{figure*}[ht!]
      \centering
      \includegraphics[width=\textwidth]{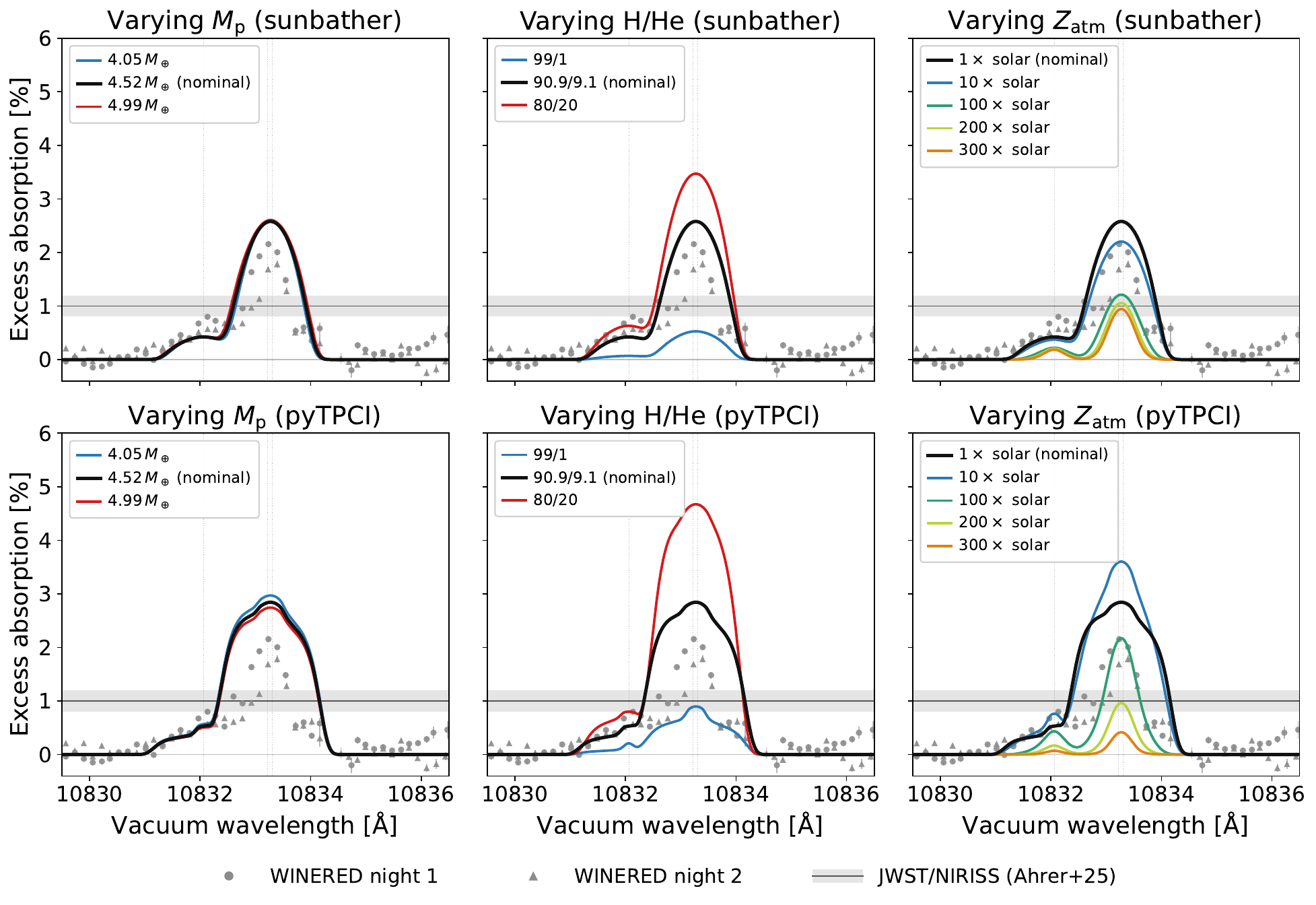}
      \caption{Models of the GJ 3090b He$^*$ signal with \textsf{sunbather} \citep{Linssen2024} in the top panels and \textsf{pyTPCI} \citep{Rosener2025} in the lower panels. Gray circles and triangles denote our first and second night of WINERED observations (binned to $R\sim68,000$), respectively, and the horizontal lines indicate the expected high-resolution excess absorption corresponding to the JWST/NIRISS measurement from \citet{Ahrer2025}.}
      \label{fig:modeling}
  \end{figure*}

\subsection{Comparison to \citet{Ahrer2025}}

Recently, \citet{Ahrer2025} reported a $5\sigma$ detection of He$^*$ in GJ 3090b at $R\sim650$ using JWST NIRISS/SOSS. These authors anticipated a shallower signal ($1.0\pm0.2\%$) at high resolution than we observed. To calculate the expected high-resolution signal, they used a Gaussian with 0.75~\AA~FWHM to model the intrinsic signal, which is somewhat narrower than what we observed with WINERED. To compare the results on more equal footing, we convolved our GP-subtracted WINERED transmission spectra with NIRISS's line-spread function and interpolated the result onto the pixel-level grid from \citet{Ahrer2025}. We find an expected NIRISS signal of $1304\pm38$~ppm for the first night and $1174\pm37$~ppm for the second night. Each observation exceeds the NIRISS observations (\citealt{Ahrer2025} find $434\pm79$~ppm) by at least 5$\sigma$, and the averaged observation exceeds the reported NIRISS measurement by 8.5$\sigma$, adding some weight to the case for He$^*$ variability. It is worth noting that the NIRISS measurement itself comes from two transits that were taken six months apart. The NIRISS signal does not appear to vary significantly across these transits, but the limited single-transit NIRISS SNR makes it difficult to assess. The night-to-night WINERED variation would have been 130~ppm if observed by NIRISS, undetectable at the achieved single-transit precisions.

\section{Outflow Modeling} \label{sec:modeling}
Next, we compared our observed He$^*$ signal to the predictions of multiple outflow models, focusing first on models in the literature. \citet{Ahrer2025} performed detailed non-isothermal modeling of the He$^*$ signal using the ATES code \citep{Caldiroli2021, Biassoni2024} and found an expected outflow signature of 3.5\% and mass-loss rate of $\dot{M} = 10^{10.1}~$g~s$^{-1}$, assuming a 90-10 H/He ratio (by number) and using a scaled MUSCLES \citep{Youngblood2016} spectrum of GJ 832 as a proxy. While \citet{Ahrer2025} state that this corresponds to an atmospheric lifetime of $\sim50$~Gyr (using $M_p/\dot{M}_p\approx50$~Gyr), the atmospheric mass (assuming it is predominantly H/He) is at most $4.3\%$ of $M_p$ for this sub-Neptune \citep{Almenara2022, Lamontagne2026}, so the true atmospheric lifetime in their models is $f_\mathrm{atm}M_p/\dot{M}_p$, only about 2~Gyr at most. The \textsf{sunset} model grid \citep{Linssen2025} predicts a He$^*$ signal of 3.76\% and a mass-loss rate of $10^{10.0}~$g~s$^{-1}$ assuming a solar H/He ratio, but using the scaled MUSCLES spectrum of GJ 176 as an XUV proxy instead of GJ 832. Both models agree well on the expected He$^*$ absorption signal despite differences in methodology, but both models also significantly overpredict the signal compared to the observations.

We performed our own modeling of GJ 3090b using \textsf{pyTPCI} \citep{Rosener2025} and \textsf{sunbather} \citep{Linssen2024}. A critical difference for our approach is the XUV spectrum. We started with the scaled MUSCLES spectrum of GJ 176 as a proxy \citep[similar to][]{Linssen2025}, but GJ 3090 has additional X-ray and FUV measurements from eROSITA and GALEX, respectively, allowing us to refine the XUV spectrum to match observables. We found that the scaled spectrum of GJ 176 is 17.3$\times$ less luminous in X-rays than the observation of GJ 3090 from eROSITA ($L_\mathrm{X} = 10^{27.96}$~erg~s$^{-1}$; \citealt{Magaudda2022}) and 2.3$\times$ less luminous in the FUV than the GALEX observation of GJ 3090 ($m_\mathrm{FUV} = 22.07\pm0.26$; \citealt{Bianchi2017}), while being roughly consistent with the GALEX NUV measurement ($m_\mathrm{NUV} = 20.002\pm0.063$; \citealt{Bianchi2017}). We therefore fit for a linear inflation term on the high-energy SED shortward of the GALEX NUV bandpass and extending into the X-rays to ensure that adopted proxy spectrum matched the high-energy observations for this young, active star. 

Using the updated XUV spectrum, we tested the impact of varying the planet mass, H/He ratio, and atmospheric metallicity $Z_\mathrm{atm}$ in \textsf{sunbather} and \textsf{pyTPCI}. Details of the modeling are provided in Appendix~\ref{app:modeling}, and the results are shown in Figure~\ref{fig:modeling}. The mass uncertainties are unimportant compared to other parameter uncertainties that we tested. On the other hand, the assumed H/He makes a major difference on the observed He$^*$ signal. The observations can be fit well by a solar-metallicity atmosphere that is depleted in helium, but this inference is degenerate with the adopted $Z_\mathrm{atm}$. Especially in younger systems, H/He can change significantly with differing assumptions about the fractionating atmospheric loss and the presence of a magma ocean \citep[which could result in fractionated ingassing/outgassing dynamics based on the differing partition coefficients of H and He;][]{Gupta2025, Kobayashi2026, Tang2026}.

The planetary atmospheric metallicity $Z_\mathrm{atm}$ also makes a major difference in the He$^*$ models. The predicted He$^*$ signal collapses at high metallicities, as anticipated by previous works \citep[e.g.][]{Yoshida2025, Zhang2025, Saidel2025}. \citet{Parker2025} reports the classic degeneracy between a high-metallicity atmosphere and an aerosol-covered atmosphere for GJ 3090b, which is largely broken by the He$^*$ measurement. If the atmospheric metallicity was much greater than 100$\times$ solar, the He$^*$ signal would be unobservable (Figure~\ref{fig:modeling}) unless the atmosphere was also very helium-rich. Instead, it is more likely that the lower atmosphere of the planet is obscured by an aerosol layer. This is in some tension with \citet{Ahrer2025}, who find a much higher $Z_\mathrm{atm}$ in their self-consistent retrievals on the JWST/NIRISS observations (they find $\gtrsim100\times$solar metallicity at 3$\sigma$ for clouds at $\mu$bar pressures). However, they note that their inference relies on fitting on a largely featureless spectrum, and thus it may be affected by the model assumptions. The upper and lower atmospheric metallicities need not be the same; in a fractionating outflow, the outflow metallicity is low (as lighter species preferentially escape), whereas the lower-atmospheric metallicity can be larger. However, for GJ 3090b the mass-loss rate far exceeds the critical rate for fractionation, making this possibility unlikely \citep{Cherubim2024}. 

The mass-loss rate adopted in the \textsf{sunbather} models was $\dot{M} = 10^{11}$~g~s$^{-1}$, which is in concordance with all of the self-consistently computed mass-loss rates in the \textsf{pyTPCI} models in Table~\ref{tab:pytpci} (see Appendix~\ref{app:modeling}). Only the highest-metallicity \textsf{pyTPCI} models we tested have lower mass-loss rates, down to $\dot{M} = 10^{10.5}$~g~s$^{-1}$ at $Z_\mathrm{atm} = 300\times$solar, but these models also significantly underpredict the signal. The increased mass-loss rate meaningfully changes the interpretation of GJ 3090b's evolutionary state. With a planetary mass of $4.5M_\Earth$ and a maximum H/He inventory of roughly 4.3\% \citep{Almenara2022, Lamontagne2026}, GJ 3090b is set to exhaust its H/He envelope on a timescale of $f_\mathrm{env}M_p/\dot{M}_p\sim300$~Myr. This timescale is two orders of magnitude shorter than the $M_p/\dot{M}_p\sim50$~Gyr quoted in \citet{Ahrer2025} and repeated by \citet{Lamontagne2026}, which is due to our updated XUV spectrum for the host star and our consideration of the small envelope mass fraction.

\section{Conclusions}
\label{sec:conclusion}
The metastable He 10833~\AA~triplet (He$^*$) is a powerful tracer of atmospheric loss processes and atmospheric composition in transiting exoplanets. We initiated the WINERED Helium Consortium to survey He$^*$ in sub-Neptunes, a planetary population for which atmospheric loss processes and atmospheric composition are both matters of debate. Our goal is to survey He$^*$ across a broad statistical sample (aiming for $\gtrsim30$ sub-Neptunes) at high precision (aiming for 0.2\% precision). As part of the Consortium, we observed two consecutive transits of the low-density sub-Neptune GJ 3090b and found clear evidence for a strong planetary outflow. We detected a peak absorption of $2.033\pm0.086\%$ (23.6$\sigma$) on the first night (during which we also observed a stellar flare) and $1.619\pm0.122\%$ (13.3$\sigma$) on the second night. Our high-SNR detections on each night demonstrate that Magellan II/WINERED is capable of precise He$^*$ constraints on the smallest exoplanets.

The scale of the outflow is remarkable for a small exoplanet: the helium triplet is opaque out past 50\% of the planetary Roche radius, indicating a vigorous outflow. The He$^*$ signal is not Doppler-shifted, and also varied slightly (2.8$\sigma$) between our two observations. Our measured absorptions are 8.5$\sigma$ deeper than the recent NIRISS He$^*$ measurement from \citet{Ahrer2025}, further supporting the case for variability. GJ 3090b therefore adds to the growing sample of variable He$^*$ absorption signals \citep{ZhangHD189, Guilluy2024, Levine2024, Allart2025, Bennett2026, Cherubim2026, Gressier2026, Radica2026}, although additional data would be helpful in confirming this finding.

The width of the feature is broad ($25$~km~s$^{-1}$), consistent with a hot photoevaporative outflow rather than a cooler core-powered wind, as predicted for a planet of this size, mass, and equilibrium temperature by \citet{Misener2026}. All sub-Neptune He$^*$ detections thus far exhibit similar broadening \citep{Zhang2023a, Orell-Miquel2024, Zhang2025}. However, \citet{Ballabio2025} caution that this may result from detection bias: they argue that core-powered flows may be relatively poor environments for populating the metastable state. More detailed simulations of He$^*$ absorption in more realistic \citep[i.e. non-isothermal,][]{Misener2025} core-powered winds, coupled with the large statistical sample of outflow measurements that we will collect, may help to place stronger constraints on the role of core-powered mass loss across the sub-Neptune population.

We made one-dimensional hydrodynamical simulations of the outflow using \textsf{pyTPCI} \citep{Rosener2025} and \textsf{sunbather} \citep{Linssen2024}. When making the simulations, we took into account X-ray constraints from eROSITA \citep{Magaudda2022} and both FUV and NUV constraints from GALEX \citep{Bianchi2017} to more accurately capture the high-energy SED for this young, active star. Our models demonstrate that the atmospheric metallicity is unlikely to greatly exceed $\sim100\times$ solar. This breaks the classic metallicity-aerosol degeneracy for GJ 3090b: non-detections of molecular features in this planet's lower atmosphere \citep{Ahrer2025, Parker2025} are more likely due to aerosol coverage than high metallicity.

The hydrodynamical simulations also suggest that the mass-loss rate is about an order of magnitude larger than previously anticipated for this system, on the order of $10^{11}$~g~s$^{-1}$. Combined with the fact that the H$_2$/He inventory of this planet is at most only 4.3\% \citep{Almenara2022, Lamontagne2026}, we find that GJ 3090b has an envelope loss timescale of $f_\mathrm{env}M_p/\dot{M}_p\sim$300~Myr. This is two orders of magnitude shorter than the $50$~Gyr timescale quoted by \citet{Ahrer2025}, which they calculated using $M_p/\dot{M}_p$ (i.e. not taking the small envelope mass fraction into account). We are likely witnessing a consequential moment in GJ 3090b's evolutionary history, where photoevaporation is starting to significantly deplete the planet's light element inventory \citep{Cherubim2025}. Future Magellan II/WINERED observations of other targets will help to unveil the nature and timescale of this upper atmospheric transition across the sub-Neptune population.

\begin{acknowledgments}
We acknowledge the WINERED instrument team members who supported these observations at Las Campanas Observatory, including Meghan O'Brien, Shogo Otsubo, Hiroaki Sameshima, and Tomomi Takeuchi. We further acknowledge the contributions of Yuki Sarugaku, Noriyuki Matsunaga, Yuji Ikeda, Daisuke Taniguchi, and others that have contributed to the successful operations of WINERED at Las Campanas Observatory. We thank James Owen, Hilke Schlichting, Eva-Maria Ahrer, Romain Allart, Vincent Bourrier, Dany Mounzer, and Yann Carteret for helpful conversations. We also thank the staff members at Las Campanas Observatory who assisted with telescope operations, including Roger Leiton and Carla Fuentes. We used Claude Opus 5 for assistance with coding. 

Z. L. acknowledges support from the McDonnell Center for the Space Sciences at Washington University in St. Louis. 
C. C. was supported by NASA through the NASA Hubble Fellowship grant HST-HF2-51601.001-A, awarded by the Space Telescope Science Institute, which is operated by the Association of Universities for Research in Astronomy, Inc., for NASA, under contract NAS5-26555.
M.L-M. is supported by individual research time under NASA contracts NAS5-26555 and NAS5-03127 to the Associated Universities for Research in Astronomy for the operation of the Hubble Space Telescope and James Webb Space Telescope Science Operations Centers at STScI.
A. McW. and J. T. acknowledge a Carnegie Venture Grant, which funded the installation and fabrication of ancillary equipment to enable WINERED at Las Campanas Observatory.

\facilities{Magellan:Clay (WINERED), ADS, NASA Exoplanet Archive}
\end{acknowledgments}

\clearpage

\appendix
\section{The Stellar Flare} \label{app:flare}
\begin{figure}[h!]
    \centering
    \includegraphics[width=0.7\linewidth]{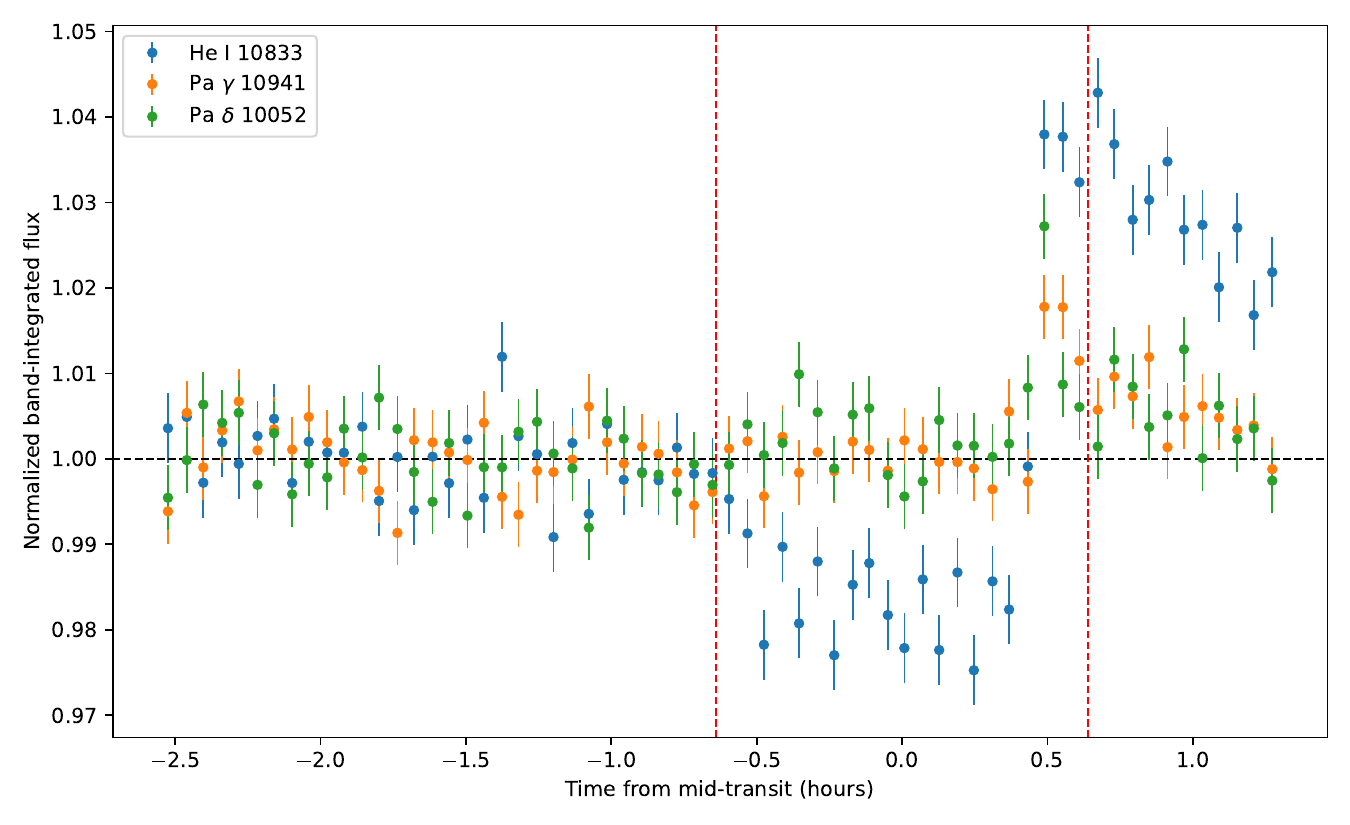}
    \caption{Spectrophotometric light curves for the first night of observations on GJ 3090b in He$^*$ (blue), Paschen $\gamma$ (orange), and Paschen $\delta$ (green). The dashed red lines indicate the predicted start and end of optical transit.}
    \label{fig:flare}
\end{figure}

During the first transit of GJ 3090b, the helium triplet brightened suddenly in the stellar rest frame just before egress. There were no stoppages or changes in strategy for this observation at any point, and the weather conditions were also quite stable throughout the first night. A natural astrophysical explanation is a stellar flare. Flares are not uncommon in He$^*$ observations of M stars \citep[e.g.][]{Fuhrmeister2020, Kanodia2022}, and GJ 3090 is known to be active with $\log(R'_\mathrm{HK}) = -4.416\pm0.048$ \citep{Almenara2022} and X-ray luminosity $L_\mathrm{X} = 10^{27.96}$~erg~s$^{-1}$ \citep{Magaudda2022}. To confirm that this event was indeed a flare, we also investigated the Paschen $\gamma$ and $\delta$ lines, which are covered by WINERED HIRES-Y in orders 161 and 176, respectively. Both of these lines also trace flares \citep[e.g.][]{Fuhrmeister2023}. We reduced the data in these orders similarly to the metastable helium order, and we integrated all three lines in a $\pm0.5$\AA~window around each of the line positions in the stellar rest frame for each spectrum. The resulting spectrophotometric light curves (normalized by the pre-ingress baseline) are shown in Figure~\ref{fig:flare}. All three tracers show an impulsive brightening followed by a slow decay, confirming that this is indeed a stellar flare.

\section{Modeling Details} \label{app:modeling}
We modeled the outflow of GJ 3090b using \textsf{sunbather} \citep{Linssen2024} and \textsf{pyTPCI} \citep{Rosener2025}. Both models have their individual shortcomings, which we hoped to overcome by combining multiple approaches. Briefly, \textsf{sunbather} couples the NLTE photoionization code \textsf{Cloudy} \citep{Ferland1998, Ferland2017, Chatzikos2023} with the 1D Parker wind code \textsf{p-winds} \citep{Dossantos2022}. This model assumes an initial isothermal Parker wind (with an assumed mass-loss rate and thermosphere temperature) calculated with \textsf{p-winds} and then determines the non-isothermal temperature structure using \textsf{Cloudy}, iterating until the temperature structure is converged. This approach converges relatively quickly, but outputs depend sensitively on the assumed Parker wind properties. On the other hand, \textsf{pyTPCI} is a self-consistent outflow modeling approach based on The \textsf{PLUTO}-\textsf{Cloudy} Interface from \citet{Salz2015}. Here, the hydrodynamics code \textsf{PLUTO} self-consistently treats the outflow launching using heating and cooling rates calculated with \textsf{Cloudy}. We take the outflow density, velocity, temperature and He$^*$ profiles from \textsf{pyTPCI} and perform radiative transfer in \textsf{p-winds} (in an equivalent method to the \textsf{sunbather} radiative transfer approach) to produce model He$^*$ spectra. Although the wind-launching process is treated more realistically by \textsf{pyTPCI}, the run time is much slower, and convergence is a challenge especially in high-metallicity outflows \citep{Rosener2025}. 

All of our models used the planetary, stellar, and orbital parameters from \citet{Lamontagne2026} along with the scaled XUV spectrum of GJ 176 described in the main text. We assessed the influence of three different uncertain model parameters: the planet mass $M_p$, the hydrogen-to-helium number ratio H/He, and the planetary atmospheric metallicity $Z_\mathrm{atm}$. Our nominal model has the mass fixed at the \citet{Lamontagne2026} value and H/He and $Z_\mathrm{atm}$ fixed to solar. We then re-ran the models at $\pm1\sigma$ on $M_p$, at H/He ratios of 99/1 and 80/20, and $Z_\mathrm{atm}$ of 10, 100, 200, and 300$\times$ solar. Due to the differing model strategies, each model also requires some individual assumptions. For \textsf{sunbather}, we assume an initial isothermal Parker wind at $\dot{M} = 10^{11}$~g~s$^{-1}$ \citep[based on an \textsf{ATES} model run;][]{Caldiroli2021} and an initial temperature of $T_0 = 4500$~K. This is roughly consistent with what is observed for $\dot{M}$ in the self-consistent \textsf{pyTPCI} models. For \textsf{pyTPCI} we adopted a lower-boundary number density of $10^{14}$~cm$^{-3}$. We adopt the same scheme as \citet{Rosener2025} where we run the model for at least $t = 100$~flow times (all numbers are in flow time units of $R_p/(1$~km~s$^{-1})$) before turning on advection, after which we attempt to run to $t = 1000$ flow times (but stop short if the metastable helium number density stops changing by more than 5\%). For the slow metal-rich cases, we instead aim to run the model for $t = 5$~flow times before turning on advection, after which we run to $t = 50$ (again stopping short if the metastable helium number density is converged). For the higher-metallicity runs, the initial timestep used was much smaller. Model parameters and outputs for the \textsf{pyTPCI} runs are summarized in Table~\ref{tab:pytpci}. 

\begin{figure}
    \centering
    \includegraphics[width=\linewidth]{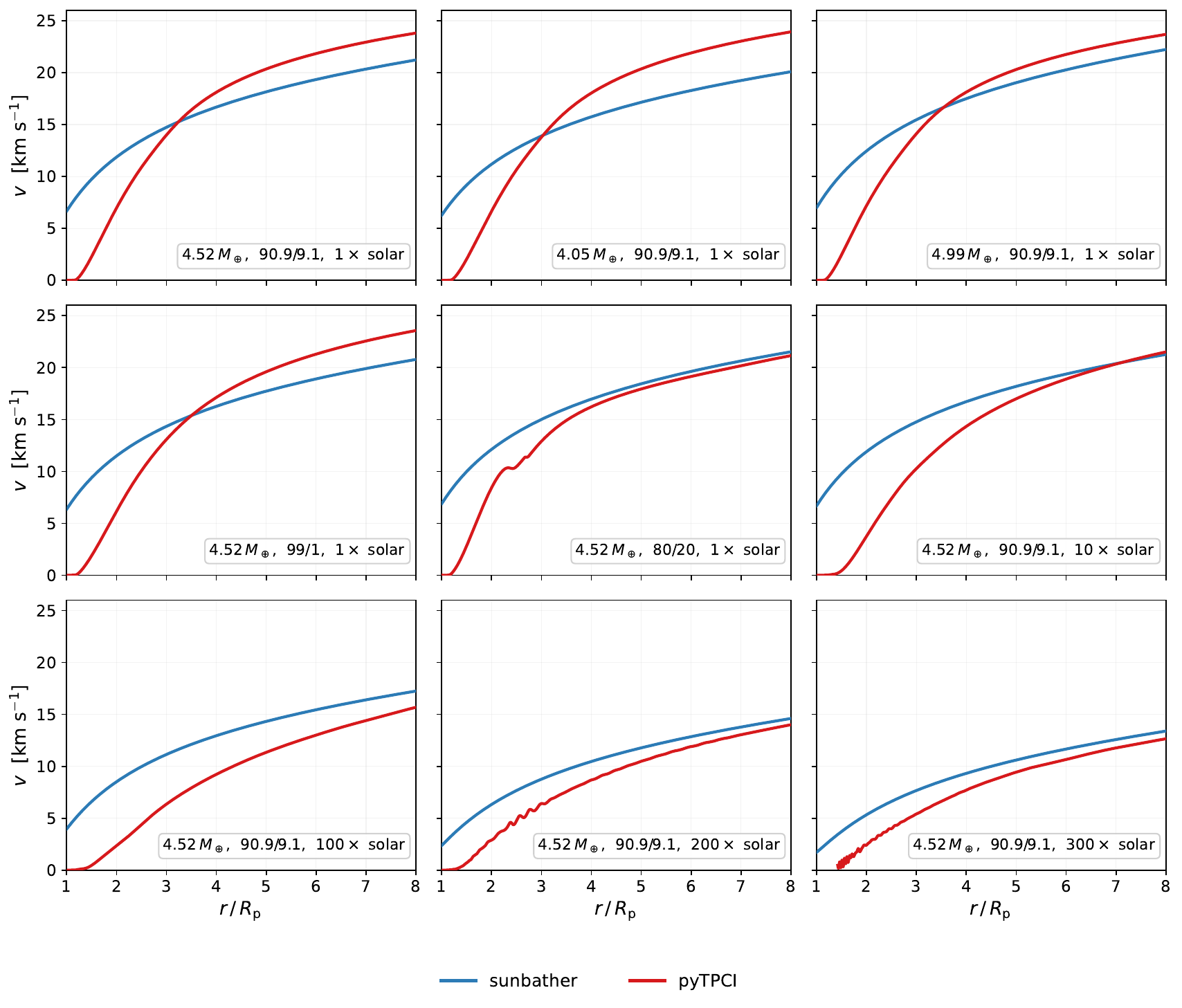}
    \caption{Velocity profiles for the \textsf{sunbather} and \textsf{pyTPCI} models.}
    \label{fig:v}
\end{figure}

\begin{figure}
    \centering
    \includegraphics[width=\linewidth]{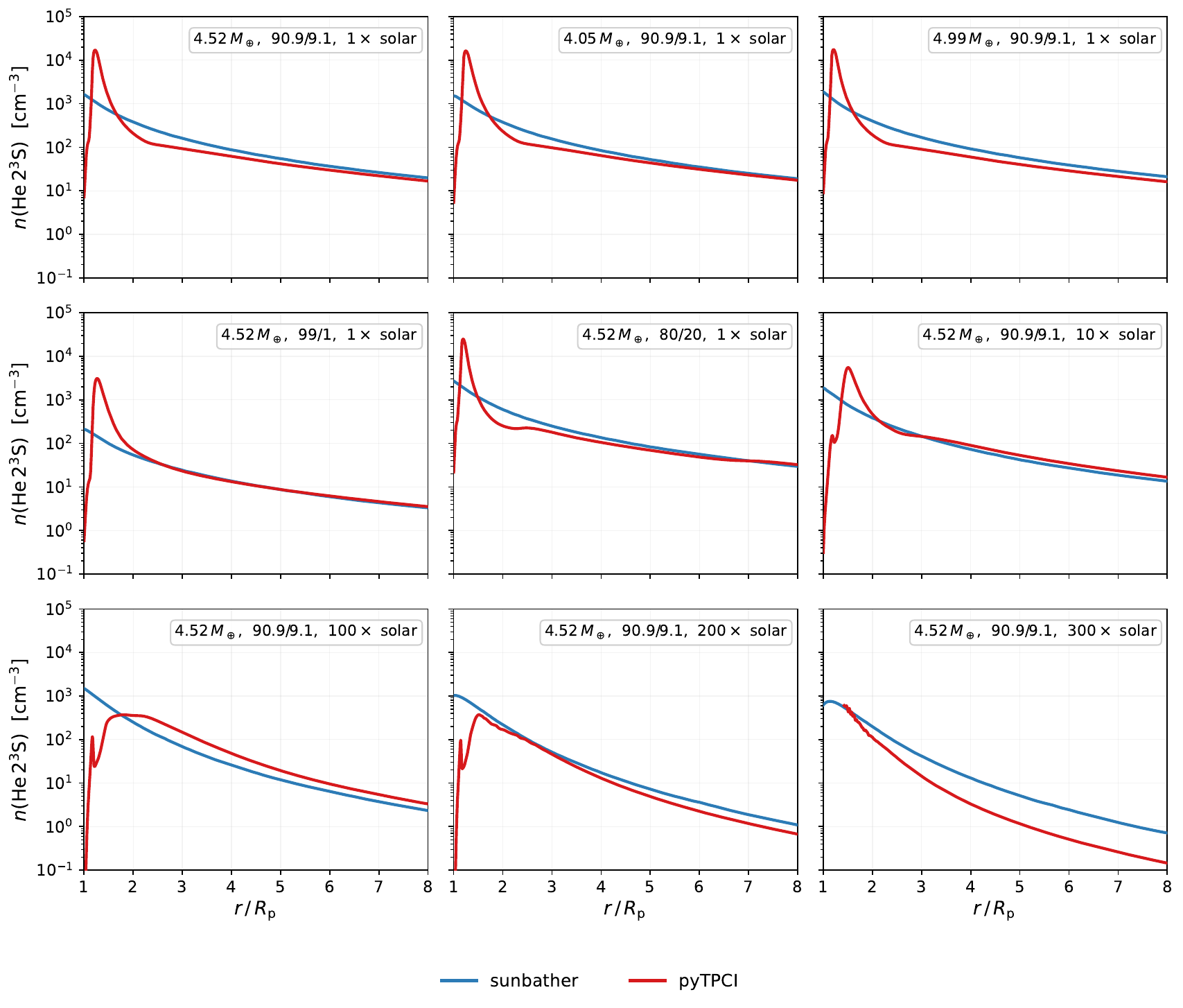}
    \caption{Metastable helium number density profiles for the \textsf{sunbather} and \textsf{pyTPCI} models.}
    \label{fig:nhe}
\end{figure}

The predicted He$^*$ signatures for all models are shown in Figure~\ref{fig:modeling}, and we show the model velocity and metastable helium number density profiles in Figure~\ref{fig:v} and Figure~\ref{fig:nhe}, respectively. The helium triplet behaves fairly similarly between the two models. In neither model does the mass uncertainty significantly affect the model, whereas H/He makes a major difference across both grids. The nominal models also anticipate a stronger He$^*$ signature than the high-$Z_\mathrm{atm}$ models in all cases. One difference between the models is that as $Z_\mathrm{atm}$ increases, He$^*$ drops monotonically in \textsf{sunbather}, whereas it increases and then decreases in \textsf{pyTPCI}. This is because the \textsf{sunbather} models are pinned to the initial $\dot{M}$ assumption, whereas the \textsf{pyTPCI} models are self-consistently including the heating and process into the wind-launching. At modest $Z_\mathrm{atm}\sim10\times$ solar the metal heating results in a stronger $\dot{M}$, whereas at $Z_\mathrm{atm}\gtrsim100\times$ solar, metal cooling results in a decreased $\dot{M}$ \citep{Zhang2025}. Another key difference between the models is the predicted broadening, which arises from how the flow behaves at the lower boundary. In Figure~\ref{fig:nhe}, lower-metallicity \textsf{pyTPCI} models show a persistent helium ``bump'' close to R$_p$, while the Parker wind models are smooth. The bump occurs before the flow is fully accelerated, which leads to the predicted He$^*$ signatures from \textsf{pyTPCI} having multiple velocity components. Although the multicomponent signature appears to be a worse fit to the data than the smoother signature predicted by \textsf{sunbather}, higher resolving power \citep[e.g.][]{Leung2025} is necessary to assess these differences carefully.

\begin{deluxetable*}{lcccccccc}[t]
\tablewidth{0pt}
\tablecaption{Summary of \textsf{pyTPCI} Models for GJ 3090 b \label{tab:pytpci}}
\tablehead{
\colhead{Model} & \colhead{$M_\mathrm{p}$} & \colhead{H/He} & \colhead{$Z_\mathrm{atm}$} &
\colhead{$t_\mathrm{pre}$} & \colhead{$t_\mathrm{post}$} &
\colhead{$\Delta t_\mathrm{init}$} &
\colhead{$\log \dot{M}$} & \colhead{$A_\mathrm{peak}$} \\
\colhead{} & \colhead{$M_\oplus$} & \colhead{by number} & \colhead{$\times$ solar} &
\colhead{} & \colhead{} & \colhead{} &
\colhead{g\,s$^{-1}$} & \colhead{\%}}
\startdata
Nominal & 4.52 & 90.9/9.1 & 1 & 251 & 1000 & 0.1 & 11.06 & 2.84 \\
$-1\sigma$ mass & 4.05 & 90.9/9.1 & 1 & 251 & 1000 & 0.1 & 11.09 & 2.97 \\
$+1\sigma$ mass & 4.99 & 90.9/9.1 & 1 & 300 & 1001 & 0.1 &  11.04 & 2.74 \\
Helium poor & 4.52 & 99/1 & 1 & 251 & 1000 & 0.1 & 11.15 & 0.90 \\
Helium rich & 4.52 & 80/20 & 1 & 172 & 172 & 0.1  & 11.02 & 4.67 \\
$10\times$ solar & 4.52 & 90.9/9.1 & 10 & 151 & 190 & 0.01 & 11.24 & 3.60 \\
$100\times$ solar & 4.52 & 90.9/9.1 & 100 & 7 & 50 & 0.001 & 11.08 & 2.17 \\
$200\times$ solar & 4.52 & 90.9/9.1 & 200 & 5 & 5 & 0.001 & 10.82 & 0.96 \\
$300\times$ solar & 4.52 & 90.9/9.1 & 300 & 6 & 6 & 0.001 & 10.53 & 0.42
\enddata 
\tablecomments{All times are in flow time units of $R_p/(1$~km~s$^{-1})$. $t_\mathrm{pre}$ indicates the number of flow times before advection is turned on, and $t_\mathrm{post}$ indicates the number of flow times run after advection was turned on.}
\end{deluxetable*}

\clearpage

\bibliography{references}{}
\bibliographystyle{aasjournalv7}

\end{document}